# Strong influence of non-magnetic ligands on the momentum dependent spin splitting in antiferromagnets

Lin-Ding Yuan[1], Zhi Wang[1,2*], Jun-Wie Luo[2] and Alex Zunger[1*]

[1]Energy Institute, University of Colorado, Boulder, CO 80309, USA

[2]State Key Laboratory for Superlattices and Microstructures, Institute of Semiconductors, Chinese Academy of Sciences, Beijing 100083, China

Recent studies have shown that the non-relativistic antiferromagnetic ordering could generate momentum-dependent spin splitting analogous to the Rashba effect, but free from the requirement of relativistic spin-orbit coupling. Whereas the classification of such compounds can be illustrated by different spin-splitting prototypes (SSTs) from symmetry analysis and density functional theory calculations, the significant variation in bonding and structure of these diverse compounds representing different SSTs clouds the issue of how much of the variation in spin splitting can be traced back to the symmetry-defined characteristics, rather to the underlining chemical and structural diversity. The alternative model Hamiltonian approaches do not confront the issues of chemical and structural complexity, but often consider only the magnetic sublattice, dealing with the all-important effects of the non-magnetic ligands via renormalizing the interactions between the magnetic sites. To this end we constructed a 'DFT model Hamiltonian' that allows us to study SSTs at approximate 'constant chemistry', while retaining the realistic atomic scale structure including ligands. This is accomplished by using a single, universal magnetic skeletal lattice ($Ni^{2+}$ ions in Rocksalt NiO) and designing small displacements of the non-magnetic (oxygen) sublattice which produce, by design, the different SSTs magnetic symmetries. We show that (i) even similar crystal structures having very similar band structures can lead to contrasting behavior of spin splitting vs. momentum, and (ii) even subtle deformations of the non-magnetic ligand sublattice could cause a giant spin splitting in AFM-induced SST. This is a paradigm shift relative to the convention of modeling magnets without considering the non-magnetic ligand that mediate indirect magnetic interaction (e.g., super exchange).

## I. INTRODUCTION

The splitting of electronic states into spin bands has long been the focus of spin electronics [1-3] in either of the two approaches: (1) Obtaining spin splitting as a result of the Zeeman effect via applied external magnetic fields or via the internal magnetic field of a ferromagnet [4]; (2) attaining momentum dependent spin splitting in non-centrosymmetric systems via spin-orbit coupling (SOC) such as the Rashba [5] and Dresselhaus [6] effects. Recently, a third mechanism of spin splitting was proposed [7], combining the advantages of the former two approaches: It was pointed out that (3) some antiferromagnets (AFM) that satisfy certain magnetic symmetries as derived in Ref [7] offer a SOC-independent (non-relativistic) route to create momentum-dependent spin splitting and spin polarization, which do not rely on heavy-element compounds and could exist even in centrosymmetric crystals. Such spin splitting and spin polarization effect arise from the spontaneous inhomogeneous magnetic field $h(r)$ of its underlining antiferromagnetic ordering that has the same periodicity with the lattice.[8] Large AFM-induced spin splitting effect has been illustrated [7] via density functional theory (DFT) calculations in tetragonal AFM $MnF_2$. In Ref. [9] we developed a general picture of how different magnetic space groups (MSG) can lead to seven distinct "spin splitting prototypes" (SST). The spin splitting of magnetic compound is related to its MSG as well as the presence or absence of *ΘIT* symmetry -- a combination of time reversal *Θ*, spatial inversion *I*, and translational symmetries *T*. Using these derived symmetry rules, we identified [9] several compounds that are predicted to have AFM-induced spin splitting effects even without SOC (i.e., non-relativistic spin splitting), including both centrosymmetric and non-centrosymmetric crystals. However, the compounds calculated by DFT have significantly different chemical bonding and structures, possibly clouding the issue of how one can distinguish the contributions to spin splitting and spin polarization emerging from the underlying, pure symmetry-defined characteristics, from the behavior reflecting the vast chemical and structural diversity in the group. The alternative model Hamiltonian approaches used in Ref. [10-12] do not confront the issues of chemical and structural complexity, but often consider only the magnetic sublattice, dealing with the all-important effects of the non-magnetic ligands via renormalizing the interactions between the magnetic sites. Therefore, our current approach to study different spin splitting types at 'constant chemistry' is to fix the chemical identity and geometry of the magnetic ions (here, nickel), and apply subtle positional changes to the non-magnetic ligands (here, oxygen) so that the net increase in total energy is small (i.e., less than 30 meV/atom). This corresponds effectively to a 'DFT model Hamiltonian' that allows us to study spin splitting prototypes at 'constant chemistry', while retaining the realistic atomic scale structure including ligands.

Our DFT results shows that (1) similar structures of NiO having different magnetic symmetries would enable or disable different mechanisms of spin splitting (no spin splitting, Zeeman splitting, AFM-induced splitting, and SOC-induced splitting); (2) the profile of spin splitting vs. momentum is determined not only by the position of the magnetic ions, but also by the positions of the non-magnetic ligands that mediate the indirect interactions between magnetic atoms, such as super exchange[13], and double exchange[14]. Significantly, a giant non-relativistic AFM-induced spin splitting (~200 meV) emerges by a very small deform (~0.04 Å) from its no spin splitting ground state position. These insights on the role of non-magnetic ligands are consequential for fundamental understanding of magnetism [15,16] because of the large number of magnetic compounds outside elemental magnets and their intermetallic alloys that owe their chemical stability to the existence of bridging ligands.

## II. MODELING DIFFERENT SPIN SPLITTING PROTOTYPES BY APPLYING SMALL DEFORMATIONS TO THE ROCKSALT AFM NIO

### A. The unperturbed ground state rock-salt NiO is an antiferromagnetic structure with no spin splitting

NiO is a wide-gap antiferromagnetic insulator below the Néel temperature (523K) [17]. The spontaneous antiferromagnetic ordering in NiO originates from the super exchange interaction between Ni via the non-magnetic oxygen p-orbitals, as proposed by Anderson [13]. Such antiferromagnetic order creates inequivalence between the Ni atoms of neighboring (111) layers (denoted as $Ni_\alpha$ and $Ni_\beta$ in Fig. 1(a)) and hence doubles the unit cell (two formula units per magnetic unit cell). Satoshi et. al. synthesized the crystal using Bernoulli method and determined the crystal structure via X-ray diffraction, achieving a good fit (R factor= 1.6%)[18] to the cubic structure (Pearson symbol: cF8) at room temperature with lattice constant a=4.178 Å[18]. Below the Néel temperature, the cubic crystal is slightly contracted along the <111> direction to a rhombohedral unit cell[17], a deformation ignored here as it changes neither the magnetic space group nor the spin splitting prototype. The antiferromagnetic order in NiO measured by neutron diffraction [19] shows that the crystal has ferromagnetic (111) sheets that couple antiferromagnetically with magnetic moment perpendicular to the propagation vector $k = \left(\frac{1}{2}, \frac{1}{2}, \frac{1}{2}\right)$. We adopt the experimental [11-2] direction magnetization for the DFT calculations).The magnetic moments is aligned parallel or antiparallel to the $[11\bar{2}]$ direction[20]. Symmetry analysis (given in Appendix A) shows that reorienting the magnetization may lead to a change in magnetic space group but does not change its MSG type or the spin splitting prototypes. This means that even with the spin polarized (collinear) setting, assuming the default [001] direction of magnetization, DFT would still be able to predict the correct spin splitting behaviors of NiO.

B. **Gentle deformations of the NiO ground state structure converting it into other spin splitting prototypes**

Our previous work[7,9] revealed that the spin splitting of magnetic compound is related to its MSG type as well as the presence or not of *ΘIT* symmetry. Formally, it can be summarized into two symmetry design principles (DP's) for non-relativistic spin polarization and spin splitting: (1) violation of *ΘIT* symmetries, and (2) MSG being type I or type III. Depending on which of the DP's (DP-1 and/or DP-2) are satisfied/violated, one can identify seven different spin splitting prototypes[9]. The classification includes four AFM prototypes plus one ferromagnetic and two non-magnetic prototypes.

To model all the seven prototypes in the same stichometry, we use the ground state NiO as the base structure, then search for small deformations that could tune the ground state NiO into the other 6 different spin splitting prototypes[9]. Such deformations, as shown in Fig. 1, include (i) atomic displacements off the rock-salt Wyckoff positions, and (ii) changes in the magnetic order (AFM, FM, and NM*)* via manipulations of local moments (by flipping or removing the spin moment). Note that our approach is to avoid the generation of traditional model Hamiltonian (corresponding generally to omitting and including interaction terms at will, i.e., truncation) but rather to subtly manipulate atomic positions in a crystal that realize new symmetries, and then act on these new structures by (untruncated) DFT, thus establishing their magnetic and spin properties.

*SST-1 structure—centrosymmetric MSG type III with ΘIT symmetry, showing no spin splitting*: these are AFM compounds that violate DP-1 but satisfy DP-2. The violation of DP-1 then ensures no spin splitting for both SOC off and SOC on cases. The SST-1 NiO structure can be achieved by shifting the central Ni atom ($Ni_\beta$) off center by 2$d$ (where $d$ denotes the amplitude of displacement) and the two O atoms by $d$, all along the [111] direction (see Fig. 1(b)). The resulting SST-1 structure has centrosymmetric parent space group R-3m, and magnetic space group of C2/m' (MSG type III with $\theta IT$ symmetry).

*SST-2 structure – centrosymmetric MSG type IV with ΘIT symmetry, showing no spin splitting:* these are AFM compounds that violates both DP-1 and DP-2. The ground state rock-salt AFM NiO (see Fig. 1(a)) is an example of SST-2 which has a centrosymmetric parent space group Fm-3m, and magnetic space group of $C_c$2/c (MSG type IV with *ΘIT* symmetry).

*SST-3 structure—MSG type IV without ΘIT symmetry, showing SOC-induced spin splitting in the presence of AFM:* these are AFM compounds that satisfy DP-1 but violate DP-2. This allows spin splitting only when SOC is turned on but with AFM magnetization in the background. Such magnetic background breaks the time reversal symmetry and allows for spin splitting at TRIMs (see Appendix B for examples and detailed discussion). The SST-3 NiO structure can be achieved by shifting both O atoms along [111] by the same distance $d$ while keeping the magnetic sublattices of Ni untouched (see Fig. 1(c)). The resulting SST-3 structure has non-centrosymmetric parent space group of R3m, and magnetic space group of $C_c$/c (MSG type IV without $\theta IT$).

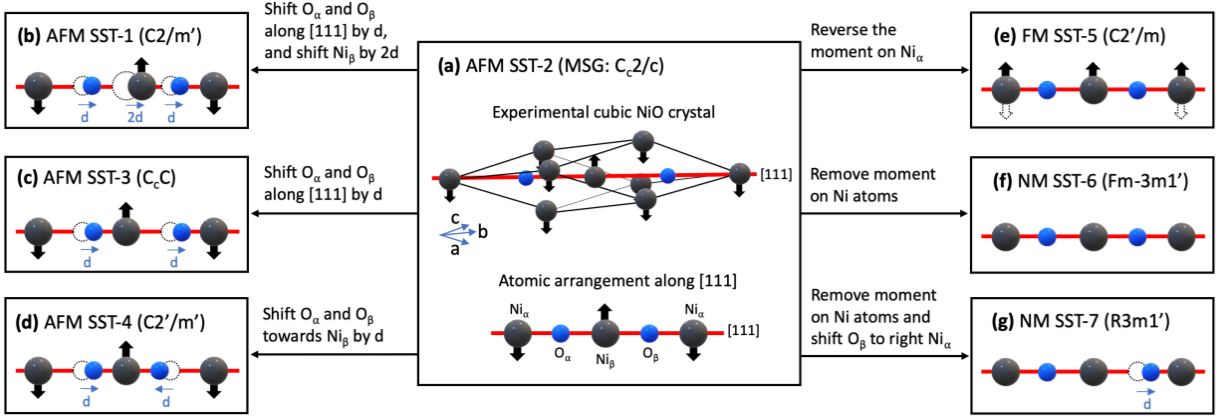

**Figure 1 | Schematic diagrams of NiO atomic and magnetic structures for the seven spin splitting prototypes**. The models are uniquely represented by the atomic arrangement along the [111] direction (indicated by red line) and the magnetic ordering on Ni (indicated by black arrows) in the unit cell. Here, we have used the observed magnetization direction $[11\bar{2}]$. $Ni_\alpha$ and $Ni_\beta$ (black circles) are the two Ni sites and $O_\alpha$ and $O_\beta$ (blue circles) are the two oxygen sites. Solid and open circles indicate, respectively, atomic positions before and after displacements *d*. **(a)** The experimental observed AFM cubic phase of NiO belong to spin splitting prototype 2 (SST-2). The other 6 prototypes **(b-g)** are derived from the AFM rock-salt NiO experimental magnetic structure (AFM SST-2) by local distortions and/or flips of magnetic moments.

*SST-4 structure—MSG type III without ΘIT symmetry, showing AFM-induced spin splitting even without SOC:* these are AFM compounds obeying both DP-1 and DP-2. Such spin splitting arises from AFM ordering (non-relativistic effect); thus, it can exist even with zero net magnetization and even when SOC is absent. The non-relativistic origin of spin splitting present in SST-4 does not rely on the unstable high-Z elements that are required for large relativistic SOC. The SST-4 NiO structure can be achieved by shifting the two oxygen atoms *towards* $Ni_\beta$ by *d*, while keeping the magnetic sublattices of Ni untouched (see Fig. 1(d)). The resulting SST-4 structure has centrosymmetric parent space group of R-3m, and magnetic space group of C2'/m' (MSG type III without $\theta IT$).

*SST-5 structure—ferromagnetic structure showing Zeeman type spin splitting:* these are ferromagnetic compounds that violate both DP-1 and DP-2. The resulting spin splitting is attributed to the Zeeman effect induced by the spontaneous net magnetization of the ferromagnet. The SST-5 NiO structure can be obtained by aligning the opposite and compensating magnetic moments of SST-2 on alternating (111) layers to the same direction, as shown in Fig. 1(e). The emergent non-zero net magnetization in this FM structure will then give rise to a Zeeman effect and split the spin up and spin down bands. The resulting FM SST-5 structure has centrosymmetric parent space group of Fm-3m, and magnetic space group of C2'/m (MSG type III without $\theta IT$).

*SST-6 structure—centrosymmetric non-magnets without spin splitting:* these are non-magnetic compounds that obey both DP-1 and DP-2. Like the CS AFM SST-1 and SST-2, the violate of DP-1 in SST-6 compounds then guarantees no spin splitting. The SST-6 NiO structure can be obtained by removing the local magnetic moments on all Ni (see Fig. 1(f)). The resulting SST-6 structure has centrosymmetric parent space group of Fm-3m, and magnetic space group of Fm-3m1' (MSG type II with $\theta IT$).

*SST-7 structure—non-centrosymmetric non-magnets, showing SOC-induced spin splitting as the Rashba-Dresselhaus effect:* these are non-magnetic structures satisfying DP-1 but violating DP-2. Such spin splitting effect occurs only when SOC is turned on, usually known as Rashba[5] and Dresselhaus[6] effect. The SST-7 NiO structure can be obtained from SST-6 by further displacing one O atom along [111] which breaks the inversion symmetry (see Fig. 1(g)). The resulting SST-7 structure has centrosymmetric parent space group of R3m, and magnetic space group of R3m1' (MSG type II with $\theta IT$).

## III. ELECTRONIC STRUCTURE OF DIFFERENT SPIN SPLITTING PROTOTYPES

To simulate the electronic properties of the NiO structures we employed DFT with the exchange correlation functional of Perdew-Burke-Ernzerhof (PBE) [21,22]. The Dudarev method [23] is used assigning an effective *U*= 4.6 eV [24] to account the on-site Coulomb repulsion energy between Ni-3d electrons. Detailed computational setting have been given in Appendix C.

### A. Electronic property of undeformed rocksalt NiO (SST-2)

For ground state NiO, we used the undeformed structure with experimental lattice constant 4.178 Å [18] and antiferromagnetic ordering aligning along $[11\bar{2}]$ direction [25]. The calculated magnetic moment on $Ni^{2+}$ ($3d^8$) is $1.7\mu_B$ comparable to the neutron-scattering results of 1.9 $\mu_B$ [26]. Electronic structure calculations show a direct gap at L of 3.55 eV and a smaller indirect gap of 2.98 eV between valence band maximum (VBM) at L and conduction band minimum (CBM) at some point on the Γ-K path (about 55% distant from Γ), which are smaller than the gap (4.3 eV) measured by combined photoemission/inverse photoemission[27].

B. Electronic properties of the seven different spin splitting prototypes of NiO structures

The DFT calculated electronic structures of the seven spin splitting prototypes presented by NiO is depicted in Fig. 1 and summarized in Table I. It compares the parent space groups, the magnetic space groups, the magnetic moment vectors (a, a, -2a) on Ni, and the band gaps, in cases of with and without the SOC term in the Hamiltonian. For all NiO structures, the displacement parameter *d* is chosen to be 0.042 Å. Different sets of displacement parameter *d* do not change the underlining magnetic space group symmetries but will has effect on the resulting spin splitting, which will also be discussed later in this section.

We note the changes in total energies relative to the ground state SST-2 are very small for the four AFM spin splitting prototypes (SST-1 to SST-4), because they differ with each other only by small displacements on oxygens: The resulting total DFT energies listed in Table I are given with respect to the undeformed rock-salt NiO (SST-2) with SOC, showing rather small destabilizations (less than 20 meV/formula unit). Removing SOC contributes to an additional small destabilization of approximately 11 meV. Moreover, the energy gaps and the magnetic moments of the four different AFM SST's are similar (see Table I row 6,7) because (i) the atomic distortion are kept small, and (ii) the SOC strength in NiO is negligible since its constituent elements Ni (Z=28) and O (Z=16) are both rather low-Z elements. The other type of deformation (i.e., change of the magnetic order) would result in remarkable variations in properties. Specifically, the change from AFM (SST-2) to FM (SST-5) leads to a total energy increase by more than 100 meV/formula unit, a smaller band gap (1.67 eV), and a slightly larger magnetic moment on Ni sites (1.8 μB); while the change from AFM to NM (centrosymmetric SST-6 and non-centrosymmetric SST-7) leads to an enormous increase in energy (~1700 meV/formula unit), zero gap, and zero magnetic moment on Ni.

**Table I | Summary of DFT calculated total energies, band gaps, magnetic moments and maximal spin splitting with and without SOC for the seven NiO structures belonging to the seven spin splitting prototypes (SST) (Fig.1) together with their space groups and magnetic space groups**. Noncollinear settings of magnetic moment oriented along $[11\bar{2}]$ are considered for both SOC off and SOC on. Maximum spin splitting is given for

the top four valence bands and the bottom four conduction bands. For FM case SST-5, because of the difficulty in identifying pairs of spin splitting states in the heavily entangled band structure, we provide spin splitting values only by the order of magnitude (>1eV). DFT results are obtained using with Perdew-Burke-Ernzerhof (PBE) exchange correlation functional with U=4.6 eV. Different structures are deformed from experimental structure of NiO SST-2 with displacement parameter *d*=0.042 Å as shown in Figure 1.

| NiO structure prototypes: | | AFM SST-1 | AFM SST-2 | AFM SST-3 | AFM SST-4 | FM SST-5 | NM SST-6 | NM SST-7 |
|---|---|---|---|---|---|---|---|---|
| Spin splitting consequence: | | No spin splitting | No spin splitting | SOC | AFM | Zeeman | No spin splitting | Rashba-Dresselhaus |
| Parent space group: | | R-3m | Fm-3m | R3m | R-3m | Fm-3m | Fm-3m | R3m |
| Magnetic space group: | | C2/m' | $C_c$2/c | $C_c$C | C2'/m' | C2'/m | Fm-3m1' | R3m1' |
| Total energy (meV/fu) | SOC off | 31 | 11 | 16 | 26 | 134 | 1678 | 1592 |
| | SOC on | 20 | 0 | 5 | 15 | 112 | 1662 | 1683 |
| Magnetic moment (a, a, -2a) ($\mu B$) | SOC off | 0.682 | 0.682 | 0.682 | 0.682 | 0.723 | 0 | 0 |
| | SOC on | 0.681 | 0.681 | 0.681 | 0.681 | 0.721 | 0 | 0 |
| Band gap (eV) | SOC off | 2.97 | 3.00 | 3.00 | 2.72 | 1.67 | Gapless | Gapless |
| | SOC on | 2.96 | 3.0 | 3.00 | 2.72 | 1.67 | Gapless | Gapless |
| Max SS for top four VB (meV) | SOC off | 0.5 | 0 | 0 | 403.1 | >1eV | 0 | 0 |
| | SOC on | 0.7 | 0 | 60.6 | 403.1 | >1eV | 0.0 | 45.0 |
| Max SS for lowest four CB (meV) | SOC off | 0.3 | 0 | 0 | 285.0 | >1eV | 0 | 0 |
| | SOC on | 0.4 | 0 | 128.8 | 285.5 | >1eV | 0.0 | 48.1 |

Despite the similarities in electronic and magnetic properties of AFM spin splitting prototypes (SST-1 to SST-4), their spin splitting consequences are distinct and differ greatly in the ensuring splitting amplitude. The last 2 rows of Table I give the maximum spin splitting for the top four valence bands (denoted as VB1, VB2, VB3, and VB4 in decreasing order of band energy) and the bottom four conduction bands (denoted as CB1, CB2, CB3, and CB4 in increasing order of band energy) with and without SOC. We note three groups of spin splitting consequences, which are consistent [9] with the SST classifications based on symmetry. (a) **No spin splitting**: AFM SST-1 and AFM SST-2, and centrosymmetric NM SST-6, all have zero spin splitting throughout the Brillouin zone for both with and without SOC (the small non-zero values below 1 meV of SST-1 are regarded as numerical errors in DFT calculations). (b) **SOC-induced spin splitting**: AFM SST-3 and non-centrosymmetric NM SST-7, both have non-zero spin splitting only when SOC is included in the Hamiltonian, and such spin splitting become zero when SOC is excluded, which therefore is referred as SOC-induced spin splitting. (c) **AFM-induced spin splitting**: AFM SST-4 has a large momentum-dependent but SOC-unrelated spin splitting referred to as AFM-induced.

Note again that the only difference among SST-2, SST-3, and SST-4 NiO structures is the oxygen displacement (Fig. 1), yet they show fundamentally different consequences on the spin splitting. Moreover, the AFM-induced spin splitting in SST-4 can take remarkable amplitudes of 403.1 meV (for VB1-4), much larger than the SOC-induced splitting in SST-3 (60.6 meV for VB1-4) or the conventional Rashba-Dresselhaus SOC-induced spin splitting in SST-7 (45 meV for VB1-4), the latter two as the common splitting amplitude in a weak SOC material (here, NiO).

To reveal the distinct physical origins of the SOC-induced spin splitting structure (SST-3) and the AFM-induced spin splitting structure (SST-4), we introduce a scaling factor $\lambda_{soc}$ ($0 < \lambda_{SOC} < 1$) to the SO Hamiltonian term $H_{SOC} = \frac{\hbar}{2m_e^2 c^2} \frac{K(r)}{r} \frac{dV(r)}{dr} \hat{L} \cdot \hat{S}$ in the DFT formalism[28], where $\hat{L} = \hat{r} \times \hat{p}$ is the orbital angular momentum operator, $\hat{S}$ is the spin operator, $V(r)$ is the spherical part of the effective all-electron potential within the projector augmented plane wave (PAW) sphere, and $K(r) = \left(1 - \frac{V(r)}{2m_e c^2}\right)^{-2}$. By controlling the scaling factor $\lambda_{soc}$, we are then able to tune the strength of the SOC.

Figure 2 shows the spin splitting of spin pair 1 (VB1 and the next valence state with opposite spin polarization, which can be VB2, VB3 or lower valence band) as a function of the scaling factor $\lambda_{soc}$ for both SST-3 and SST-4. The identification of the spin pairs sometimes is not straightforward as the assignment of members of spin pair may change from one k point to another due to possible band crossing and anti-crossing. In practice, for each k point, we search for spin states with opposite sign of spin polarization (projecting on magnetization direction $[11\bar{2}]$) from VB1 downwards and CB1 upwards (see Appendix D for details of how to identify spin pairs). As evidenced in Figure 2(a), the SOC-induced spin splitting of SST-3 AFM NiO at both W and U have a relativistic origin (linear to $\lambda_{soc}$), and become constantly zero at Γ point as enforced by the persisting $\theta T$ symmetry in SST-3. In contrast, as shown in Figure 2(b), the AFM-induced spin splitting in SST-4 NiO is an order of magnitude larger than that of SST-3 (at W and Γ). Significantly, the resulting spin splitting are insensitive to SOC strength suggesting their non-relativistic origin.

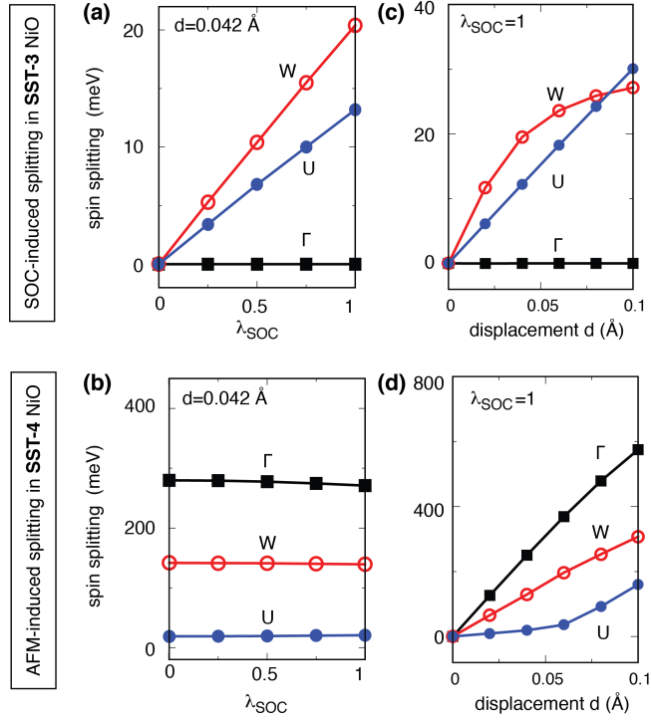

**Figure 2 | Spin splitting of valence bands (the first spin pair)** as a function of spin orbit coupling strength $\lambda_{soc}$ for (a) relativistic SOC-induced splitting in the presence of AFM (SST-3) and (b) non-relativistic AFM-induced splitting (SST-4) using a fixed displacement $d=0.042$Å along [111] direction; and as a function of displacement $d$ for (c) relativistic SOC-induced splitting in the presence of AFM (SST-3) and (b) non-relativistic AFM-induced splitting (SST-4) with fixed SOC strength $\lambda_{SOC}=1$.

We have shown how small distortions in the non-magnetic oxygen atom position results in surprisingly fundamental changes in spin splitting consequences for different SST's. To see if such dependence of spin splitting to ligands is also consistent for the same SST, we examine the amplitudes of spin splitting for the SST-3 and SST-4 NiO structures as we continuously shift the oxygen (increase in *d*).

Figure 2(c-d) illustrate the spin splitting in SST-3 and SST-4 as a function of displacement *d* for spin pair 1. Clearly, the relation of spin splitting to oxygen displacement *d* strongly depends on what SST the structure belongs to. In SST-3 (Fig. 2(c)), the dependence of SOC-induced spin splitting on *d* is weak at all three k points. The splitting vs *d* relation is quadratic at W, but linear at Γ and U. While in AFM-induced SST-4, the dependence of AFM-induced spin splitting on *d* is much stronger and rather linear at the Γ and W points while being quadratic at U. We see such non-relativistic AFM-induced splitting is very sensitive to the deformation, as a small increase of displacement (from 0.0 to 0.1 A) could result in a remarkable raise of 200-600 meV in spin splitting.

### C. Spin polarized band structure and spin splitting dispersion of individual prototypes.

As noted in Fig. 2, for each SST, the spin splitting behavior can be different at different k points. We calculate the spin splitting dispersion along certain high-symmetry k-paths for all NiO structures. We focus on the spin splitting of the top four valence bands as they are separated in energy from the other bands, thus free from the complex problem of bands entangling.

Figure 3 shows the spin polarized band structures and dispersions of spin splitting of the four AFM prototypes (SST-1 to SST-4). Each prototype has two panels, being the band structures of SOC off (left panel) and SOC on (right panel). Only one of the two sub prototypes of SST-3 (3A and 3B) and SST-4 (4A and 4B) are shown, precisely being SST-3B (non-centrosymmetric) and SST-4A (centrosymmetric)). In Fig. 3, for k-points where the spin polarizations of bands are very close to zero, the definition of spin pair and spin splitting can be uncertain; we use grey patches with no data points to indicate such rare k-paths. As discussed for Table I, the SS consequences of these AFM SST's can be separated into three groups:

*(a) Prototypes having no spin splitting either with or without SOC (SST-1, SST-2):* The bands of centrosymmetric AFM SST-1 and SST-2 without SOC show similar dispersion but differ in band crossing and anticrossing on L-U and L-K (indicated by red and green circles), whereas adding SOC enables coupling between opposite spin states therefore changes all crossing band to anticrossing as noticed on Γ-X, L-U, and L-K. The corresponding spin splitting are all zero with vanishing spin polarization as mapped by grey line color in the spin polarized band plots in Fig. 3(a-d).

*(b) Prototypes that have spin splitting only when SOC is present (SST-3):* The SOC-induced splitting AFM SST-3 without SOC shows fully degenerate bands. It shares the same topology of degeneracies with SST-2 (no spin splitting AFM prototype) for the top four valence bands, but differs from the latter in band crossing and anticrossing on L-U and L-K (indicated by circles). Adding SOC lifts the degeneracy in AFM SST-3. Such SOC-induced spin splitting has been mapped in Fig. 3(f) on high symmetry k-paths such as X-W-K and U-W-L with red/blue showing the spin-up/spin-down polarized bands. The spin splitting dispersion of $k$ in Fig. 3(g,h) shows that the such SOC-induced spin splitting (purple for the splitting of the first spin pair, and blue for the splitting of the second spin pair) in AFM SST-3 is relatively small in magnitude (<60 meV) reflecting the weak SOC strength in NiO.

*(c) Prototypes that have spin splitting regardless of SOC (SST-4):* The bands of AFM SST-4 without SOC show similar dispersion and effective mass to other AFM spin splitting prototypes SST-1 (no splitting), SST-2 (no splitting), and SST-3 (SOC-induced splitting), but show unique features of well separated branches of spin-up and spin-down bands (identified by the spin projection on the magnetization direction $[11\bar{2}]$; see the color bars on the right-hand side of Figure 3), whereas adding SOC contributes to only small changes in band structure (again, noticed as the change from crossing to anticrossing between opposite spin bands). We find that the AFM-induced spin splitting, exist even when SOC is off (see Fig. 3(i)), is ~200 meV in magnitudes, which is much larger than the SOC-induced splitting in SST-3 structure (<60 meV). Such large spin splitting is presumably attributed to the strong local magnetic moment on Ni sites.

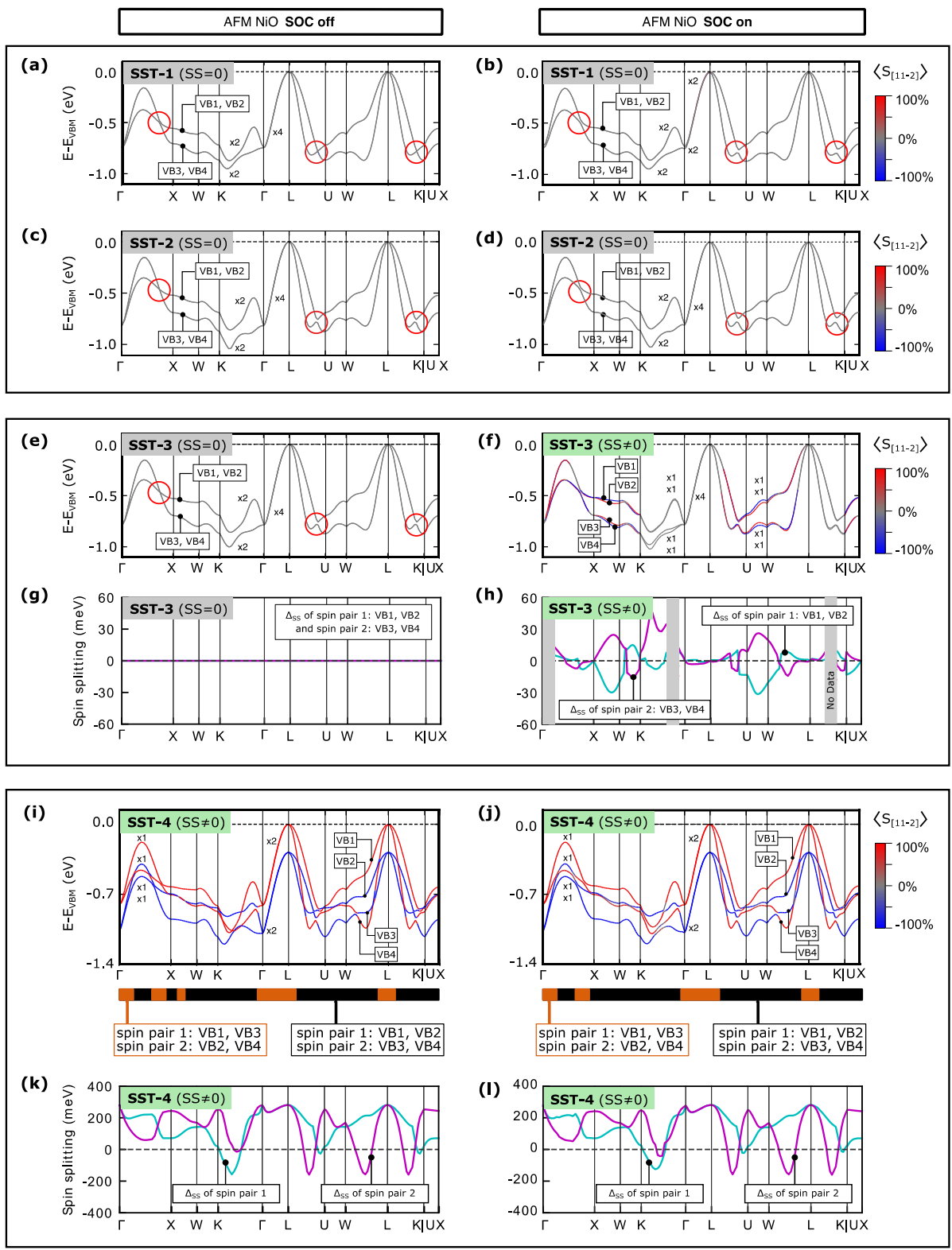

**Figure 3 | DFT calculated spin polarized band structures and spin splitting of the top four valence bands in AFM NiO prototypes (SSTs) defined in Fig.1.** (a,b) for SST-1 (no spin splitting case) with SOC off (a) and SOC on(b). (c,d)

for SST-2 (no spin splitting case) with SOC off (c) and SOC on(d); (e,f,g,h) for SST-3 (SOC-induced spin splitting case) of the four spin polarized bands with SOC off (e) and SOC on(f); and the spin splitting with SOC off (g) and SOC on(h); and (i,j,k,l) SST-4 (AFM-induced spin splitting) of the four spin polarized bands with SOC off (i) and SOC on(j); and the spin splitting with SOC off (k) and SOC on(l). The top four VB's of NiO prototypes are labeled in the order of decreasing band energy as VB1, VB2, VB3 and VB4 regardless of band crossing or anticrossing (red circles). These bands are grouped into two spin pairs (1 and 2) of neighboring bands with opposite spin polarization (see Appendix D for details). The grouping can be either spin pair 1 consisting of VB1 and VB2 and spin pair 2 consisting of VB3 and VB4, indicated by black regions in panel (k) and (l), or spin pair 1 consisting of VB1 and VB3 and spin pair 2 consisting of VB2 and VB4, indicated by orange regions in panel (k) and (l). The color bars on the right-hand side provide the spin projection on the magnetization direction $[11\bar{2}]$.

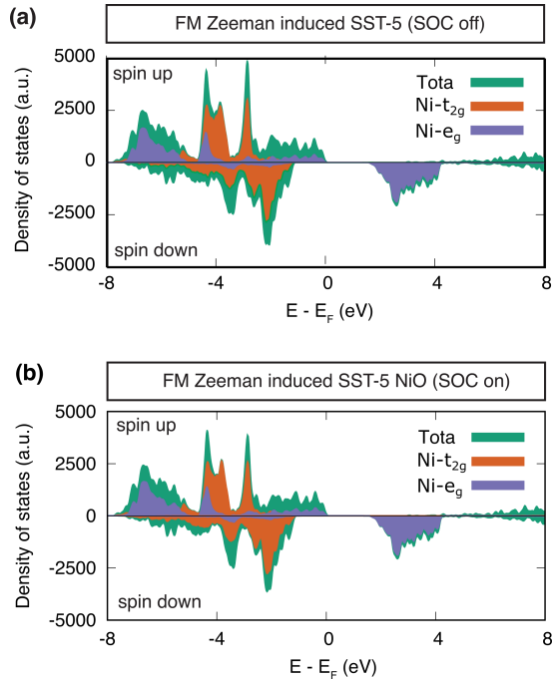

**Figure 4 | DFT results of spin and orbital projected density of states showing Zeeman splitting in FM SST-5.** The spin-projected (up and down; identified by the spin projection on the magnetization direction $[11\bar{2}]$) and orbital-projected (Ni-$t_{2g}$ and Ni-$e_g$) density of states of FM SST-5 (a) when SOC is off and (b) when SOC is on.

In addition to the four AFM SST's there are FM SST-5, NM SST-6 (centrosymmetric), and NM SST-7 (non-centrosymmetric):

*(d) Zeeman splitting in ferromagnets (SST-5):* For SST-5, the undeformed rock-salt crystal but with a FM ordering parallelly in the direction of $[11\bar{2}]$ is used for the calculation. The DFT results (Fig. 4 (a)) show a direct gap at Γ of 1.66 eV and a magnetic moment on Ni of 1.8 $\mu_B$, which is slightly larger than the AFM magnetic moment. Because of the difficulty in identifying pairs of spin splitting states in the heavily entangled band structure, the spin splitting is not evaluated to a number, but shown by the nondegenerate spin-projected density of states (DOS) of up and down spins (identified by the spin projection on the magnetization direction $[11\bar{2}]$). The separated spin-up and spin-down DOS seen in Fig. 4 for SOC off imply that the spin splitting in FM SST-5 presence even when SOC is off. Different from AFM SST-4 which also shows SOC-unrelated spin splitting, the spin splitting in FM arise from a different mechanism known as Zeeman effect due to the non-zero net magnetization, the latter of which has a larger magnitude of ~1 eV.

*(e) No spin splitting in centrosymmetric NM SST-6 and SOC-induced (Rashba/Dresselhaus) spin splitting in non-centrosymmetric NM SST-7.* Because of the removal of local magnetic moments, $Ni^{2+}$ ions ($3d^8$) will have fully occupied $t_{2g}$ orbitals (6 electrons for 6 degenerated orbitals) but only half-occupied $e_g$ orbitals (2 electrons for 4 degenerated orbitals), therefore the band gaps found in AFM and FM prototypes close to zero in non-magnetic prototypes (SST-6 and SST-7). For centrosymmetric NM SST-6, just like SST-1 and SST-2 (no spin splitting AFM prototypes), the $\theta IT$ symmetry is present and consequently ensures double spin degeneracy for every k-points. This is evidenced by the spin degenerate band with vanishing spin polarization shown in Fig. 5(a) and the corresponding zero spin splitting for spin pair 1 and 2 shown in Fig. 5(b). In non-centrosymmetric NM SST-7 (Fig. 5(c-d)), a small (<40 meV) spin splitting arises only when SOC is on. Such splitting has the same relativistic origin as the SOC-induced spin splitting in AFM SST-3, except that the latter has AFM order present.

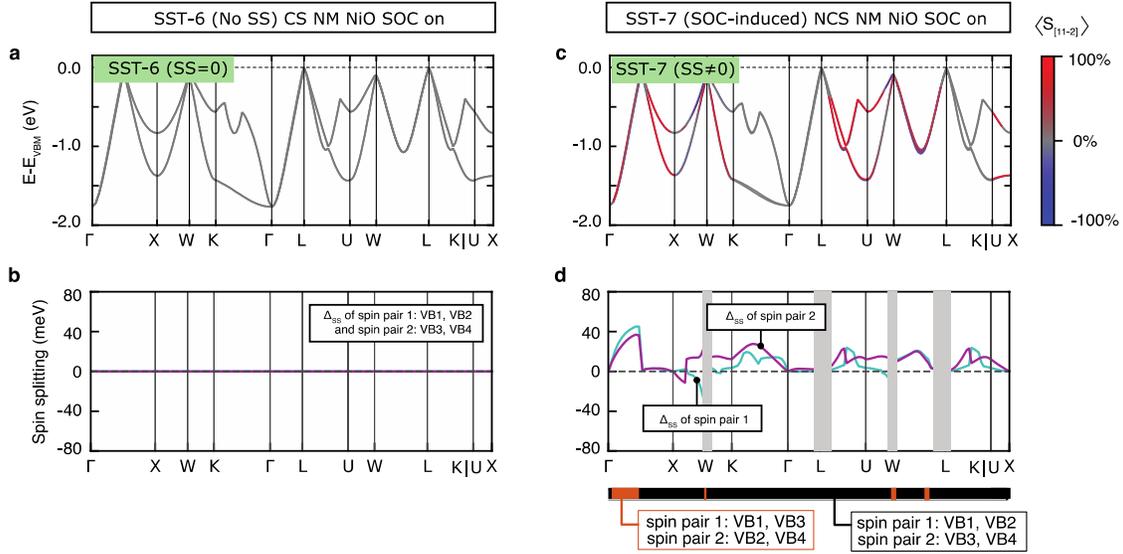

**Figure 5 | Spin polarized energy bands and spin splitting of top four valence bands in non-magnetic NiO prototypes structures.** (a) spin polarized bands and (b) spin splitting for centrosymmetric SST-6 with SOC on; (c) spin polarized bands and (d) spin splitting for non-centrosymmetric SST-7 with SOC on.

## IV. Conclusions

In this work, we focus on the role of non-magnetic ligands by examining systems of different spin splitting prototypes at "constant chemistry". In contrast with previous approaches that attempted to introduce the effect of ligands on the magnetic symmetry by scaling the contribution to amplitude of exchange interactions without the presence of ligands in the Hamiltonian, we opt for the latter. Our approach is implemented via a set of subtle structural deformation and magnetic deformations (from AFM to FM and NM) derived from one single base structure, the classic rock-salt AFM NiO[25], that each satisfies one of the seven spin splitting symmetry conditions. To address the role of the often-neglected, non-magnetic ligand atoms, the structural deformations are carefully designed to displace O atoms only while keeping Ni positions fixed. First-principles density functional theory method has been used to study these seven designed structures of NiO.

Our main findings are: (1) the profile of spin splitting vs. momentum can be greatly affected by the Wyckoff position of the non-magnetic ions occupy. Rather similar structures of NiO, sharing the same magnetic sublattices but differing only by subtle deformations of the non-magnetic ions, can have significantly different spin splitting behaviors and significant differences in splitting amplitude; (2) we studied how the amplitude of spin splitting evolves as the deformation of Oxygen in SST-3 and SST-4 NiO; we show a giant (~200 meV) non-relativistic AFM-induced spin splitting in the slightly deformed SST-4 NiO (with only a small Oxygen displacement of 0.04 Å from its original position in no spin splitting SST-2 NiO), in contrast to the small (~20 meV) SOC induced-spin splitting. We envision the important role of the non-magnetic ions as revealed in NiO structures, which is a serious warning to the tradition of considering only magnetic sublattices when modeling a magnet.

**ACKNOWLEGEMNETS**

We thank Emmanuel Rashba for many discussions that led to Ref. [7,9] and much more. The work at CU Boulder was supported by the National Science Foundation (NSF) DMR-CMMT Grant No. DMR-1724791 that supported the formal theory development of this work. The electronic structure calculations of this work were supported by the U.S. Department of Energy, Office of Science, Basic Energy Sciences, Materials Sciences and Engineering Division under Grant No. DE-SC0010467. J.-W.L. was supported by the National Natural Science Foundation of China (NSFC) under Grant No. 61888102. This work used resources of the National Energy Research Scientific Computing Center, which is supported by the Office of Science of the U.S. Department of Energy; and the Extreme Science and Engineering Discovery Environment (XSEDE) supercomputer resources.

# Appendix

**A. Compounds of different magnetization orientations can have different magnetic space groups but belong to the same MSG type and therefore the same spin splitting prototype.**

When SOC is on, the spatial rotation is also associated to the rotation of spin or local magnetic moment. The transformation of magnetic moment under spatial rotation thus depends on the magnetization direction: (1) a $\pi$ rotation around axes perpendicular to the magnetization orientation will reverse the on-site magnetic moment, but (2) a $\pi$ rotation around axes parallel to the magnetization orientation will keep the local magnetic moment unchanged. It opens the possibility for changing MSG via only changing the magnetization orientation. For example, the AFM NiO with $[11\bar{2}]$ oriented magnetic moments has MSG $C_c2/c$, while the AFM NiO with [1-10] oriented magnetic moments has a different MSG $C_c2/m$. In the former case, a $\pi$ rotation around [1-10] at center $Ni_\beta$ (denoted as $C_{2[1-10]}$) keeps the atomic arrangement in unit cell but reverses the magnetic moment on Ni atoms, therefore its combination with time reversal $\theta$ (i.e., $\theta C_{2[1-10]}$) is a symmetry of its MSG $C_c2/c$; in the latter case, $C_{2[1-10]}$ keeps the atomic arrangement in unit cell and also keeps the orientation of magnetic moment on Ni atoms unchanged, thus $C_{2[1-10]}$ (no need to combine with $\theta$) is a symmetry of MSG $C_c2/m$ of the new model.

Although the change of magnetization orientation would change the MSG of a magnetic compound, it won't change the underlining MSG type or the presence or not of $\theta IT$ symmetry. Therefore, the change of magnetization orientation will not change the SST of an AFM compound. This conclusion is not limited to the NiO system but rather general to all collinear antiferromagnetic compounds.

**B. Spin splitting at time reversal invariant moments (TRIMs)**

The k point that differs from k by a reciprocal lattice vector is known as a time reversal invariant moment (TRIM). Typically, in a 3-dimensional reciprocal space, the TRIMs are (0,0,0), (1/2,0,0), (0,1/2,0), (0,0,1/2), (1/2,1/2,0), (1/2,0,1/2), (0,1/2,1/2), and (1/2,1/2,1/2) in unit of reciprocal lattices.

In non-magnetic systems where the time reversal symmetry is preserved, the time reversal symmetry will also be a symmetry of the Hamiltonian at the TRIMs, hence guarantee a spin degeneracy and no spin splitting at the TRIMs. But unlike non-magnetic systems, the AFM compounds violate the time reversal symmetry $\theta$ thus allow spin splitting to occur at TRIM. However, this does not happen in SST-3 compounds where the MSG type is type IV, as the existence of an alternative symmetry, the $\theta T$ symmetry, will again ensure spin degeneracy at TRIMs in the same manner as $\theta$ for non-magnets. The only AFM prototype that allows TRIM spin splitting is SST-4 (AFM-induced spin splitting). Such spin splitting on TRIM is observed in the example of SST-4 NiO, where a spin splitting of approximately 200 meV are noticed (Fig. 3 (i-l)) at Γ point. The formal method of determining TRIM degeneracy in magnetic systems requires the co-representation theory and the use of Herring theorem [29] as briefly described below.

<u>Herring's theorem</u>: In AFM compounds of a given MSG $M$, the degeneracies of the electronic states are determined by the irreducible co-representation of $M$ [30,31] which can be deduced from the irreducible representation of $\boldsymbol{G}$. The dimension of the deduced co-representation is 2 times the dimension of the irreducible representation. Whether if such co-representation is irreducible (i.e., spin degeneracy) or reducible (i.e., spin splitting) can be achieved by Herring theorem [29]:

$$\sum_{B \in \{\theta R_m\}} \chi^\Delta(B^2) = \begin{cases} g \ case\ (a) \\ 0 \ case\ (b) \\ -g \ case\ (c) \end{cases} \quad (1)$$

where $\chi$ is the character of irreducible representation $\Delta$ of group $\boldsymbol{G}$ with index $g$. Case (a) co-representation is reducible, meaning that the spin states having such co-representation will split; while case (b) and (c) co-representations are both irreducible, i.e., spin degeneracy will be preserved. For example, every pair of spin state in MSG type IV AFM holds either case (b) or (c) co-representation, hence must be spin degenerate: this is because that for MSG type IV AFM $\theta T \in \{\theta R_m\}$, thus it always has $\chi^\Delta(\theta T^2) = -1$ hence cannot be case (a).

**C. DFT settings:** The band structures were calculated by density function theory (DFT) method with Perdew-Burke-Ernzerhof (PBE) exchange correlation functional[22] with U=4.6 eV and J=0 eV on Ni 3d orbitals[24] following the simplified rotationally invariant approach introduced by Dudarev et. al[23]. The atomic structures are either the experimental structure[32] or the deformed structures changing only the internal atomic positions but keeping the experimental lattice vectors. We adopt a plane-wave basis of up to 500 eV energy cutoff, a Γ-centered sampling of k-mesh $11 \times 11 \times 11$ (6000 k points in the first Brillouin zone), and the tetrahedron smearing method for the calculations of a self-consistent charge density. For all NiO models we chose the high-symmetric k-paths from the conventional rhombohedral first Brillouin zone to show the band structures.

**D. Determining the spin pairs:** The assignment of the members of a spin pair can change from one k point to another, due to band crossing and anti-crossing (see red circles in Fig. 3). To solve this issue, we label the top four valence bands in decreasing order of band energy as VB1, VB2, VB3 and VB4. At every k point, the VB1-VB4 states can be grouped into two spin pairs (spin pairs 1 and 2) of neighboring bands with opposite spin polarization projected on the magnetization direction $[11\bar{2}]$. The two spin pairs can be either (a) spin pair 1 of VB1 and VB2 and spin pair 2 of VB3 and VB4, or (b) spin pair 1 of VB1 and VB3 and spin pair 2 of VB2 and VB4. For each spin pair the spin splitting takes a positive value when the up-spin state energy is higher than the down-spin state energy, and vice versa. However, not all $k$ points have a clear definition for the spin pairs. The same procedure applies to determine spin splitting pairs for other valences bands and conduction bands.

For states where the spin polarizations are very close to zero, it is hard to determine whether the state is spin up or spin down, due to the introducing of noncollinearity by SOC and the existence of numerical errors. We use grey patches in Fig. 3 with no data points below it to indicate such rare k-paths where the above definition of spin pairs is not applicable.

**REFERENCES**

[1] S. A. Wolf, D. D. Awschalom, R. A. Buhrman, J. M. Daughton, S. v. Molnár, M. L. Roukes, A. Y. Chtchelkanova, and D. M. Treger, *Spintronics: A Spin-Based Electronics Vision for the Future*, Science **294**, 1488 (2001)

[2] I. Žutić, J. Fabian, and S. D. Sarma, *Spintronics: Fundamentals and applications*, Rev Mod Phys **76**, 323 (2004)

[3] A. Fert, *The present and the future of spintronics*, Thin Solid Films **517**, 2 (2008)

[4]  P. Zeeman, *The Effect of Magnetisation on the Nature of Light Emitted by a Substance*, Nature **55**, 347 (1897)

[5]  E. Rashba and V. Sheka, *Symmetry of energy bands in crystals of wurtzite type II. Symmetry of bands with spin-orbit interaction included*, Fiz. Tverd. Tela, Collected Papers (Leningrad) **2**, 62 (1959) English translation: https://iopscience.iop.org/1367-2630/17/5/050202/media/njp050202_suppdata.pdf

[6]  G. Dresselhaus, *Spin-Orbit Coupling Effects in Zinc Blende Structures*, Physical Review **100**, 580 (1955)

[7]  L.-D. Yuan, Z. Wang, J.-W. Luo, E. I. Rashba, and A. Zunger, *Giant momentum-dependent spin splitting in centrosymmetric low-Z antiferromagnets*, Physical Review B **102**, 014422 (2020)

[8]  S. I. Pekar and E. I. Rashba, *Combined resonance in crystals in inhomogeneous magnetic fields*, Zh. Eksperim. i Teor. Fiz. **47** (1964) English translation: Sov. Phys. - JETP 20, 1295 (1965)

[9]  L.-D. Yuan, Z. Wang, J.-W. Luo, and A. Zunger, *Prediction of low-Z collinear and noncollinear antiferromagnetic compounds having momentum-dependent spin splitting even without spin-orbit coupling*, Physical Review Materials **5**, 014409 (2021)

[10] S. Hayami, Y. Yanagi, and H. Kusunose, *Momentum-Dependent Spin Splitting by Collinear Antiferromagnetic Ordering*, J Phys Soc Jpn **88**, 123702 (2019)

[11] S. Hayami, Y. Yanagi, and H. Kusunose, *Bottom-up design of spin-split and reshaped electronic band structures in antiferromagnets without spin-orbit coupling: Procedure on the basis of augmented multipoles*, Physical Review B **102**, 144441 (2020)

[12] S. Hayami, Y. Yanagi, and H. Kusunose, *Spontaneous Antisymmetric Spin Splitting in Noncollinear Antiferromagnets without Relying on Atomic Spin-Orbit Coupling*, 2001.05630 (2020)

[13] P. W. Anderson, *Antiferromagnetism. Theory of Superexchange Interaction*, Physical Review **79**, 350 (1950)

[14] C. Zener, *Interaction between the d-Shells in the Transition Metals. II. Ferromagnetic Compounds of Manganese with Perovskite Structure*, Physical Review **82**, 403 (1951)

[15] B. D. Cullity and C. D. Graham, *Introduction to magnetic materials* (John Wiley & Sons, 2011).

[16] D. Jiles, *Introduction to magnetism and magnetic materials* (CRC press, 2015).

[17] H. Rooksby, *A note on the structure of nickel oxide at subnormal and elevated temperatures*, Acta Crystallographica **1**, 226 (1948)

[18] S. Sasaki, K. Fujino, Tak, Eacute, and Y. Uchi, *X-Ray Determination of Electron-Density Distributions in Oxides, MgO, MnO, CoO, and NiO, and Atomic Scattering Factors of their Constituent Atoms*, Proceedings of the Japan Academy, Series B **55**, 43 (1979)

[19] C. G. Shull, W. A. Strauser, and E. O. Wollan, *Neutron Diffraction by Paramagnetic and Antiferromagnetic Substances*, Physical Review **83**, 333 (1951)

[20] J. Baruchel, M. Schlenker, K. Kurosawa, and S. Saito, *Antiferromagnetic S-domains in NiO*, Philosophical Mag B **43**, 853 (1981)

[21] J. P. Perdew, K. Burke, and M. Ernzerhof, *Generalized Gradient Approximation Made Simple*, Physical Review Letters **77**, 3865 (1996)

[22] J. P. Perdew, J. A. Chevary, S. H. Vosko, K. A. Jackson, M. R. Pederson, D. J. Singh, and C. Fiolhais, *Atoms, molecules, solids, and surfaces: Applications of the generalized gradient approximation for exchange and correlation*, Physical Review B **46**, 6671 (1992)

[23] S. L. Dudarev, G. A. Botton, S. Y. Savrasov, C. J. Humphreys, and A. P. Sutton, *Electron-energy-loss spectra and the structural stability of nickel oxide: An LSDA+U study*, Physical Review B **57**, 1505 (1998)

[24] M. Cococcioni and S. de Gironcoli, *Linear response approach to the calculation of the effective interaction parameters in the LDA+U method*, Physical Review B **71**, 035105 (2005)

[25] E. Ressouche, N. Kernavanois, L.-P. Regnault, and J.-Y. Henry, *Magnetic structures of the metal monoxides NiO and CoO re-investigated by spherical neutron polarimetry*, Phys B Condens Matter **385-386**, 394 (2006)

[26] E. Ressouche, N. Kernavanois, L.-P. Regnault, and J.-Y. Henry, *Magnetic structures of the metal monoxides NiO and CoO re-investigated by spherical neutron polarimetry*, Phys B Condens Matter **385**, 394 (2006)

[27] G. A. Sawatzky and J. W. Allen, *Magnitude and Origin of the Band Gap in NiO*, Physical Review Letters **53**, 2339 (1984)

[28] S. Steiner, S. Khmelevskyi, M. Marsmann, and G. Kresse, *Calculation of the magnetic anisotropy with projected-augmented-wave methodology and the case study of disordered Fe1−xCox alloys*, Physical Review B **93**, 224425 (2016)

[29] C. Herring, *Effect of Time-Reversal Symmetry on Energy Bands of Crystals*, Physical Review **52**, 361 (1937)

[30] J. O. Dimmock and R. G. Wheeler, *Symmetry Properties of Wave Functions in Magnetic Crystals*, Physical Review **127**, 391 (1962)

[31] R.-X. Zhang and C.-X. Liu, *Topological magnetic crystalline insulators and corepresentation theory*, Physical Review B **91**, 115317 (2015)

[32] W. L. Roth, *Magnetic Structures of MnO, FeO, CoO, and NiO*, Physical Review **110**, 1333 (1958)